\documentclass[10pt]{article}

\newlength{\minitwocolumn}
\font\teneufm=eufm10
\font\seveneufm=eufm7
\font\fiveeufm=eufm5
\newfam\eufmfam
\textfont\eufmfam=\teneufm
\scriptfont\eufmfam=\seveneufm
\scriptscriptfont\eufmfam=\fiveeufm

\makeatletter
\@addtoreset{equation}{section}
\makeatother

\newtheorem{thm}{Theorem}[section]
\newtheorem{prop}[thm]{Proposition}

\newtheorem{dfn}[thm]{Definition}

\title{
\Large{\bf
THE $q$-WAKIMOTO REALIZATION OF\\
THE SUPERALGEBRAS $U_q(\widehat{sl}(N|1))$ AND $U_{q,p}(\widehat{{sl}}(N|1))$
}}
\begin{document}
\maketitle
\begin{center}
{TAKEO KOJIMA}
\\~\\
{\it
Department of Mathematics and Physics,
Graduate School of Science and Engineering,\\
Yamagata University, Jonan 4-3-16, Yonezawa 992-8510,
Japan\\
kojima@yz.yamagata-u.ac.jp}
\end{center}

~\\

\begin{abstract}
We give bosonizations of 
the superalgebras 
$U_q(\widehat{sl}(N|1))$ and $U_{q,p}(\widehat{sl}(N|1))$
for an arbitrary level $k \in {\bf C}$.
We introduce the submodule
by the $\xi$-$\eta$ system,
that we call the $q$-Wakimoto realization.
\end{abstract}

~\\

\section{Introduction}

Bosonizations are known to be a powerful method
to construct correlation functions in 
not only
conformal field theory
\cite{Bouwknegt-McCarthy-Pilch1},
but also exactly solvable lattice models \cite{Jimbo-Miwa}.
The quantum algebra $U_q(g)$ and the elliptic algebra $U_{q,p}(g)$
play an important role
in exactly solvable lattice models.
The level parameter $k$ plays an important role 
in representation theory for $U_q(g)$ and $U_{q,p}(g)$.
Bosonizations for an arbitrary level $k$
are completely different from those of level $k=1$.
In the case for level $k=1$,
bosonizations
have been constructed
for quantum algebra $U_q(g)$ 
in many cases
$g=(ADE)^{(r)}$, $(BC)^{(1)}$, $G_2^{(1)}$,
$\widehat{sl}(M|N)$, $osp(2|2)^{(2)}$
\cite{
Frenkel-Jing, Jing1,
Bernard, Jing-Koyama-Misra, Jing2,
Kimura-Shiraishi-Uchiyama, Zhang, Yang-Zhang}.
Using the dressing method developed in
non-twisted algebra \cite{Jimbo-Konno-Odake-Shiraishi}
and twisted algebra $A_2^{(2)}$ \cite{Kojima-Konno},
we have bosonizations
of the elliptic algebra $U_{q,p}(g)$ for
$g=(ADE)^{(1)}, (BC)^{(1)}, G_2^{(1)}$ and $A_2^{(2)}$.
In the case of an arbitrary level $k$,
bosonizations have been constructed
only for $U_q(\widehat{sl}(N))$ 
\cite{Matsuo, Shiraishi, Awata-Odake-Shiraishi1},
$U_q(\widehat{sl}(2|1))$ \cite{Awata-Odake-Shiraishi2},
$U_{q,p}(\widehat{sl}(N))$ 
\cite{Jimbo-Konno-Odake-Shiraishi},
and $U_{q,p}(\widehat{sl}(2|1))$ \cite{Kojima1}.
In this paper we give a bosonization of 
the quantum superalgebra $U_q(\widehat{sl}(N|1))$
for an arbitrary level $k$ \cite{Kojima2}.
Using the dressing method developed in
\cite{Kojima1},
we give a bosonization of 
the quantum superalgebra $U_{q,p}(\widehat{sl}(N|1))$
for an arbitrary level $k$.
The level $k$ bosonizations
on the boson Fock space
of $U_q(\widehat{sl}(N))$ and 
$U_q(\widehat{sl}(N|1))$
\cite{Awata-Odake-Shiraishi1, Awata-Odake-Shiraishi2, Kojima2}
are not irreducible realizations.
The construction of the irreducible
highest weight module $V(\lambda)$
is nontrivial problem.
We recall the non-quantum algebra 
$\widehat{sl}(2)$ case \cite{Bernard-Felder}. 
The irreducible highest weight module $V(\lambda)$
for the affine algebra $\widehat{sl}(2)$
was constructed from the Wakimoto realization
on the boson Fock space \cite{Wakimoto}
by the Felder complex. 
We recall the quantum algebra $U_q(\widehat{sl}(2))$ 
case \cite{Matsuo, Shiraishi, Konno}.
The irreducible highest weight module $V(\lambda)$
for $U_q(\widehat{sl}(2))$
was constructed from the level $k$ bosonizations
on the boson Fock space
\cite{Matsuo, Shiraishi}
by two steps; the first step is
the resolution by the $\xi$-$\eta$ system, and
the second step is the resolution
by the Felder complex
\cite{Konno, Bernard-Felder}.
The submodule of the quantum algebra $U_q(\widehat{sl}(2))$,
induced by the $\xi$-$\eta$ system,
plays the same role as the Wakimoto realization
of the non-quantum algebra $\widehat{sl}(2)$.
We would like to call this submodule
induced by the $\xi$-$\eta$ system
"the $q$-Wakimoto realization".
Constructions of the irreducible 
highest weight module $V(\lambda)$ 
for $U_q(\widehat{sl}(N))$ $(N\geq 3)$
and $U_q(\widehat{sl}(N|1))$ $(N\geq 2)$
are still
an open problem.
In this paper we study 
the $\xi$-$\eta$ system
and introduce the $q$-Wakimoto realization for 
the superalgebra $U_q(\widehat{sl}(N|1))$
and $U_{q,p}(\widehat{sl}(N|1))$.

This paper is organized as follows.
In section 2, after preparing notations,
we give the definition
of the quantum superalgebra $U_q(\widehat{sl}(N|1))$
and the elliptic superalgebra $U_{q,p}(\widehat{sl}(N|1))$.
In section 3 we give
bosonizations of the superalgebras
$U_q(\widehat{sl}(N|1))$ 
and $U_{q,p}(\widehat{sl}(N|1))$
for an arbitrary level $k$.
In section 4
we introduce the $q$-Wakimoto realization of 
by the $\xi$-$\eta$ system.

\section{Superalgebra $U_q(\widehat{sl}(N|1))$ and 
$U_{q,p}(\widehat{sl}(N|1))$}

In this section we recall the definitions of
the quantum superalgebra
$U_q(\widehat{sl}(N|1))$
\cite{Yamane} and
the elliptic deformed superalgebra $U_{q,p}(\widehat{sl}(N|1))$
\cite{Kojima1} for $N \geq 2$.
We fix a complex number $q \neq 0, |q|<1$.
We set
\begin{eqnarray}
~[x,y]=xy-yx,~\{x,y\}=xy+yx,~[a]_q=\frac{q^a-q^{-a}}{
q-q^{-1}}.
\end{eqnarray}
Let us fix complex numbers $r,k \in {\bf C}$,
${\rm Re}(r)>0, {\rm Re}(r-k)>0$.
We use the abbreviation $r^*=r-k$.
We set $p=q^{2r}$. 
We set the Jacobi theta functions
\begin{eqnarray}
~[u]=
q^{\frac{u^2}{r}-u}\frac{\Theta_{q^{2r}}(q^{2u})}{
(q^{2r};q^{2r})_\infty^3},
~~[u]^*=
q^{\frac{u^2}{r^*}-u}\frac{\Theta_{q^{2r^*}}(q^{2u})}{
(q^{2r^*};q^{2r^*})_\infty^3},
\end{eqnarray}
where we have used
\begin{eqnarray}
&&\Theta_p(z)=(z;p)_\infty 
(pz^{-1};p)_\infty (p;p)_\infty,~~~(z;p)_\infty
=\prod_{m=0}^\infty (1-p^{m}z).
\end{eqnarray}
The Cartan matrix 
$(A_{i,j})_{0\leq i,j \leq N}$
of the affine Lie algebra $\widehat{sl}(N|1)$ is
given by
\begin{eqnarray}
A_{i,j}=
(\nu_i+\nu_{i+1})\delta_{i,j}-
\nu_i \delta_{i,j+1}-\nu_{i+1}\delta_{i+1,j}.
\end{eqnarray}
Here we set $\nu_1=\cdots =\nu_N=+, \nu_{N+1}=\nu_0-$.

\subsection{Quantum superalgebra $U_q(\widehat{sl}(N|1))$}

In this section we recall
the definition of
the quantum affine superalgebra $U_q(\widehat{sl}(N|1))$.

\begin{dfn}~~\cite{Yamane}~
The Drinfeld generators of
the quantum
superalgebra $U_q(\widehat{sl}(N|1))$
are
\begin{eqnarray}
x_{i,m}^\pm,~h_{i,m},~c,~~(1\leq i \leq N,
m \in {\bf Z}).
\end{eqnarray}
Defining relations are
\begin{eqnarray}
&&~c : {\rm central},~[h_i,h_{j,m}]=0,\\
&&~[a_{i,m},h_{j,n}]=\frac{[A_{i,j}m]_q[cm]_q}{m}q^{-c|m|}
\delta_{m+n,0}~~(m,n\neq 0),\\
&&~[h_i,x_j^\pm(z)]=\pm A_{i,j}x_j^\pm(z),\\
&&~[h_{i,m}, x_j^+(z)]=\frac{[A_{i,j}m]_q}{m}
q^{-c|m|} z^m x_j^+(z)~~(m \neq 0),\\
&&~[h_{i,m}, x_j^-(z)]=-\frac{[A_{i,j}m]_q}{m}
z^m x_j^-(z)~~(m \neq 0),\\
&&(z_1-q^{\pm A_{i,j}}z_2)
x_i^\pm(z_1)x_j^\pm(z_2)
=
(q^{\pm A_{j,i}}z_1-z_2)
x_j^\pm(z_2)x_i^\pm(z_1)~~~{\rm for}~|A_{i,j}|\neq 0,
\\
&&
x_i^\pm(z_1)x_j^\pm(z_2)
=
x_j^\pm(z_2)x_i^\pm(z_1)~~~{\rm for}~|A_{i,j}|=0, (i,j)\neq (N,N),
\\
&&
\{x_N^\pm(z_1), x_N^\pm(z_2)\}=0,\label{def:Drinfeld8}\\
&&~[x_i^+(z_1),x_j^-(z_2)]
=\frac{\delta_{i,j}}{(q-q^{-1})z_1z_2}
\left(
\delta(q^{-c}z_1/z_2)\psi_i^+(q^{\frac{c}{2}}z_2)-
\delta(q^{c}z_1/z_2)\psi_i^-(q^{-\frac{c}{2}}z_2)
\right), \nonumber\\
&& ~~~~~{\rm for}~~(i,j) \neq (N,N),\\
&&~\{x_N^+(z_1),x_N^-(z_2)\}
=\frac{1}{(q-q^{-1})z_1z_2}
\left(
\delta(q^{-c}z_1/z_2)\psi_N^+(q^{\frac{c}{2}}z_2)-
\delta(q^{c}z_1/z_2)\psi_N^-(q^{-\frac{c}{2}}z_2)
\right), \nonumber\\
\\
&& 
\left(
x_i^\pm(z_{1})
x_i^\pm(z_{2})
x_j^\pm(z)-(q+q^{-1})
x_i^\pm(z_{1})
x_j^\pm(z)
x_i^\pm(z_{2})
+x_j^\pm(z)
x_i^\pm(z_{1})
x_i^\pm(z_{2})\right)\nonumber\\
&&+\left(z_1 \leftrightarrow z_2\right)=0
~~~{\rm for}~|A_{i,j}|=1,~i\neq N,
\end{eqnarray}
where we have used
$\delta(z)=\sum_{m \in {\bf Z}}z^m$.
Here we have 
used the abbreviation $h_i={h_{i,0}}$.
We have
set the generating function
\begin{eqnarray}
x_j^\pm(z)&=&
\sum_{m \in {\bf Z}}x_{j,m}^\pm z^{-m-1},\\
\psi_i^+(q^{\frac{c}{2}}z)&=&q^{h_i}
\exp\left(
(q-q^{-1})\sum_{m>0}h_{i,m}z^{-m}
\right),\\
\psi_i^-(q^{-\frac{c}{2}}z)&=&q^{-h_i}
\exp\left(-(q-q^{-1})\sum_{m>0}h_{i,-m}z^m\right).
\end{eqnarray}
\end{dfn}

\subsection{Elliptic Superalgebra $U_{q,p}(\widehat{sl}(N|1))$}

In this section we recall
the definition of
the elliptic superalgebra $U_{q,p}(\widehat{sl}(N|1))$.

\begin{dfn}~~~\cite{Kojima1}~~~
The elliptic superalgebra
$U_{q,p}(\widehat{sl}(N|1))$ is the associative algebra
generated by
the currents $E_j(z), F_j(z), H_j^\pm(z)$ $(1\leq j \leq N)$
and $B_{j,m} (1\leq j \leq N, m \in {\bf Z}_{\neq 0})$, 
$h_j$ $(1\leq j \leq N)$ that
satisfy the following relations.
\begin{eqnarray}
&&~[h_i,B_{j,m}]=0,~[B_{i,m},B_{j,n}]=\frac{[A_{i,j}m]_q[km]_q}{m}
\frac{[r^*m]_q}{[rm]_q}
\delta_{m+n,0},\\
&&~[h_i,E_j(z)]=A_{i,j}E_j(z),~[h_i,F_j(z)]=-A_{i,j}F_j(z),\\
&&~[B_{i,m}, E_j(z)]=\frac{[A_{i,j}m]_q}{m}
z^m E_j(z),\\
&&~[B_{i,m}, F_j(z)]=-\frac{[A_{i,j}m]_q}{m}\frac{[r^*m]_q}{
[rm]_q}
z^m F_j(z).
\end{eqnarray}
For $1\leq i,j \leq N$ such that $(i,j)\neq (N,N)$
they satisfy
\begin{eqnarray}
&&~\left[u_1-u_2-\frac{A_{i,j}}{2}\right]^*
E_i(z_1)E_j(z_2)
=
\left[u_1-u_2+\frac{A_{i,j}}{2}\right]^*
E_j(z_2)E_i(z_1),
\\
&&~\left[u_1-u_2+\frac{A_{i,j}}{2}\right]
F_i(z_1)F_j(z_2)
=
\left[u_1-u_2-\frac{A_{i,j}}{2}\right]
F_j(z_2)F_i(z_1),
\\
&&~[E_i(z_1),F_j(z_2)]
=\frac{\delta_{i,j}}{(q-q^{-1})z_1z_2}
\left(
\delta(q^{-k}z_1/z_2)H_i(q^{r}z_2)-
\delta(q^{k}z_1/z_2)H_i(q^{-r}z_2)
\right),\\
&&\{E_N(z_1),E_N(z_2)\}=0,~\{F_N(z_1),F_N(z_2)\}=0,\\
&&\{E_N(z_1),F_N(z_2)\}
=\frac{1}{(q-q^{-1})z_1z_2}
\left(
\delta(q^{-k}z_1/z_2)H_N(q^{r}z_2)-
\delta(q^{k}z_1/z_2)H_N(q^{-r}z_2)
\right).\nonumber\\
\end{eqnarray}
For $1\leq i,j \leq N$ they satisfy
\begin{eqnarray}
&&H_i(z_1)H_j(z_2)=\frac{
[u_2-u_1-\frac{A_{i,j}}{2}]^*
[u_2-u_1+\frac{A_{i,j}}{2}]}{
[u_2-u_1+\frac{A_{i,j}}{2}]^*
[u_2-u_1-\frac{A_{i,j}}{2}]}
H_j(z_2)H_i(z_1),\\
&&H_i(z_1)E_j(z_2)=
\frac{
[u_1-u_2+\frac{r^*}{2}+\frac{A_{i,j}}{2}]^*}{
[u_1-u_2+\frac{r^*}{2}-\frac{A_{i,j}}{2}]^*}
E_j(z_2)H_i(z_1),\\
&&H_i(z_1)F_j(z_2)=
\frac{
[u_1-u_2+\frac{r}{2}+\frac{A_{i,j}}{2}]}{
[u_1-u_2+\frac{r}{2}-\frac{A_{i,j}}{2}]}
F_j(z_2)H_i(z_1).
\end{eqnarray}
For $1\leq i,j \leq N$,$(i\neq N)$ such that $|A_{i,j}|=1$,
they satisfy the Serre relations.
\begin{eqnarray}
&&
~~\left\{
E_i(z_1)E_i(z_2)E_j(z)
\frac{
\left(q^{2r^*+A_{i,j}}\frac{z}{z_1};q^{2r^*}\right)_\infty 
\left(q^{2r^*+A_{i,j}}\frac{z}{z_2};q^{2r^*}\right)_\infty}{
\left(q^{2r^*-A_{i,j}}\frac{z}{z_1};q^{2r^*}\right)_\infty
\left(q^{2r^*-A_{i,j}}\frac{z}{z_2};q^{2r^*}\right)_\infty}
\left(\frac{z}{z_2}\right)^{\frac{1}{r^*}A_{i,j}}
\right.\nonumber\\
&&-(q+q^{-1})
E_i(z_1)E_j(z)E_i(z_2)
\frac{
\left(q^{2r^*+A_{i,j}}\frac{z}{z_1};q^{2r^*}\right)_\infty 
\left(q^{2r^*+A_{i,j}}\frac{z_2}{z};q^{2r^*}\right)_\infty}{
\left(q^{2r^*-A_{i,j}}\frac{z}{z_1};q^{2r^*}\right)_\infty 
\left(q^{2r^*-A_{i,j}}\frac{z_2}{z};q^{2r^*}\right)_\infty}
\nonumber
\\
&&\left.
+E_j(z)E_i(z_1)E_i(z_2)
\frac{
\left(q^{2r^*+A_{i,j}}\frac{z_1}{z};q^{2r^*}\right)_\infty
\left(q^{2r^*+A_{i,j}}\frac{z_2}{z};q^{2r^*}\right)_\infty}{
\left(q^{2r^*-A_{i,j}}\frac{z_1}{z};q^{2r^*}\right)_\infty 
\left(q^{2r^*-A_{i,j}}\frac{z_2}{z};q^{2r^*}\right)_\infty}
\left(\frac{z_1}{z}\right)^{\frac{1}{r^*}A_{i,j}}
\right\}\nonumber\\
&&\times
\frac{\left(q^{2r^*+A_{i,i}}\frac{z_2}{z_1}; q^{2r^*}\right)_\infty}
{\left(q^{2r^*-A_{i,i}}\frac{z_2}{z_1};q^{2r^*}\right)_\infty}
z_1^{-\frac{1}{r^*}(A_{i,i}+A_{i,j})}+(z_1 \leftrightarrow z_2)=0,
\end{eqnarray}
\begin{eqnarray}
&&~~\left\{F_i(z_1)F_i(z_2)F_j(z)
\frac{
\left(q^{2r-A_{i,j}}\frac{z}{z_1};q^{2r}\right)_\infty 
\left(q^{2r-A_{i,j}}\frac{z}{z_2};q^{2r}\right)_\infty}{
\left(q^{2r+A_{i,j}}\frac{z}{z_1};q^{2r}\right)_\infty 
\left(q^{2r+A_{i,j}}\frac{z}{z_2};q^{2r}\right)_\infty}
\left(\frac{z_2}{z}\right)^{\frac{1}{r}A_{i,j}}
\right.\nonumber\\
&&-(q+q^{-1})
F_i(z_1)F_j(z)F_i(z_2)
\frac{
\left(q^{2r-A_{i,j}}\frac{z}{z_1};q^{2r}\right)_\infty 
\left(q^{2r-A_{i,j}}\frac{z_2}{z};q^{2r}\right)_\infty}{
\left(q^{2r+A_{i,j}}\frac{z}{z_1};q^{2r}\right)_\infty 
\left(q^{2r+A_{i,j}}\frac{z_2}{z};q^{2r}\right)_\infty}
\nonumber
\\
&&\left.
+
F_j(z)F_i(z_1)F_i(z_2)
\frac{
\left(q^{2r-A_{i,j}}\frac{z_1}{z};q^{2r}\right)_\infty 
\left(q^{2r-A_{i,j}}\frac{z_2}{z};q^{2r}\right)_\infty}{
\left(q^{2r+A_{i,j}}\frac{z_1}{z};q^{2r}\right)_\infty 
\left(q^{2r+A_{i,j}}\frac{z_2}{z};q^{2r}\right)_\infty}
\left(\frac{z}{z_1}\right)^{\frac{1}{r}A_{i,j}}
\right\}\nonumber\\
&&
\times
\frac{\left(q^{2r-A_{i,i}}\frac{z_2}{z_1};q^{2r}\right)_\infty}{
\left(q^{2r+A_{i,i}}\frac{z_2}{z_1};q^{2r}\right)_\infty}
z_1^{\frac{1}{r}(A_{i,i}+A_{i,j})}
+(z_1 \leftrightarrow z_2)=0.
\end{eqnarray}
Here we have used $z_j=q^{2u_j}$.
\end{dfn}

\section{Bosonization}

In this section we give bosonizations of the superalgebras 
$U_q(\widehat{sl}(N|1))$ and
$U_{q,p}(\widehat{sl}(N|1))$ for an arbitrary level $k$
\cite{Awata-Odake-Shiraishi2, Kojima1, Kojima2}.

\subsection{Boson}

We fix the level $c=k \in {\bf C}$.
We introduce the bosons
and the zero-mode operators
$a_m^j, Q_a^j$ $(m \in {\bf Z},
1\leq j \leq N)$, 
$b_m^{i,j}, Q_b^{i,j}$
$(m \in {\bf Z}, 1\leq i<j \leq N+1)$,
$c_m^{i,j}, Q_c^{i,j}$
$(m \in {\bf Z}, 1\leq i<j \leq N)$.
The bosons $a_m^i, b_m^{i,j}, c_m^{i,j}$, 
$(m \in {\bf Z}_{\neq 0})$ and
the zero-mode operators $a_0^i,Q_a^i$, $b_0^{i,j},Q_b^{i,j}$,
$c_0^{i,j}, Q_c^{i,j}$ satisfy
\begin{eqnarray}
&&~[a_m^i,a_n^j]=\frac{[(k+N-1)m]_q[A_{i,j}m]_q}{m}
\delta_{m+n,0},~~~[a_0^i, Q_a^j]=(k+N-1)A_{i,j},
\\
&&~[b_m^{i,j},b_n^{i',j'}]=
-\nu_i \nu_j \frac{[m]_q^2}{m}
\delta_{i,i'}\delta_{j,j'}\delta_{m+n,0},~~~
[b_0^{i,j},Q_b^{i',j'}]=
-\nu_i \nu_j \delta_{i,i'}\delta_{j,j'},
\\
&&~[c_m^{i,j},c_n^{i',j'}]=
\frac{[m]_q^2}{m}
\delta_{i,i'}\delta_{j,j'}
\delta_{m+n,0},~~~
[c_0^{i,j},Q_c^{i',j'}]=
\delta_{i,i'}\delta_{j,j'}.
\end{eqnarray}
We impose the cocycle condition on 
the zero-mode operator $Q_{b}^{i,j}$, $(1\leq i<j \leq N+1)$ by
\begin{eqnarray}
~[Q_b^{i,j},Q_b^{i',j'}]=\delta_{j,N+1}\delta_{j',N+1}
\pi \sqrt{-1}~~~~~{\rm for}~(i,j) \neq (i',j').
\end{eqnarray}
We have the following (anti) commutation relations
\begin{eqnarray}
&&
\left[e^{Q_b^{i,j}},e^{Q_b^{i',j'}}\right]=0
~~~
(1\leq i<j \leq N, 1\leq i'<j' \leq N),\\
&&\left\{e^{Q_b^{i,N+1}},e^{Q_b^{j,N+1}}\right\}=0~~~
(1\leq i \neq j \leq N).
\end{eqnarray}
We use the standard symbol of the normal orderings $: :$.
In what follows we use the abbreviations 
$b^{i,j}(z), c^{i,j}(z), b_\pm^{i,j}(z),
a^j_\pm(z)$ given by
\begin{eqnarray}
&&b^{i,j}(z)=
-\sum_{m \neq 0}\frac{b_m^{i,j}}{[m]_q}z^{-m}+Q_b^{i,j}+b_0^{i,j}{\rm log}z,~
c^{i,j}(z)=
-\sum_{m \neq 0}\frac{c_m^{i,j}}{[m]_q}z^{-m}+Q_c^{i,j}+c_0^{i,j}{\rm log}z,
\\
&&b_\pm^{i,j}(z)=\pm (q-q^{-1})\sum_{\pm m>0}b_m^{i,j} 
z^{-m} \pm b_0^{i,j}{\rm log}q,~
a_\pm^{j}(z)=\pm (q-q^{-1})\sum_{\pm m>0}a_m^{j} 
z^{-m}\pm a_0^j {\rm log}q.
\end{eqnarray}

\subsection{Quantum Superalgebra $U_q(\widehat{sl}(N|1))$}

In this section
we give a bosonization of the quantum 
superalgebra $U_q(\widehat{sl}(N|1))$ for an arbitrary level $k$.

\begin{thm}~~\cite{Kojima2}~~
The Drinfeld currents 
$x_i^\pm(z)$,
$\psi_i^\pm(z)$, $(1\leq i \leq N)$
of $U_q(\widehat{sl}(N|1))$ 
for an arbitrary level $k$ are realized by
the bosonic operators as follows.

\begin{eqnarray}
x_i^+(z)&=&\frac{1}{(q-q^{-1})z}\sum_{j=1}^i
:\exp\left((b+c)^{j,i}(q^{j-1}z)+\sum_{l=1}^{j-1}
(b_+^{l,i+1}(q^{l-1}z)-b_+^{l,i}(q^lz))\right)
\times
\\
&\times&
\left\{
\exp\left(b_+^{j,i+1}(q^{j-1}z)-
(b+c)^{j,i+1}(q^jz)\right)-
\exp\left(b_-^{j,i+1}(q^{j-1}z)-
(b+c)^{j,i+1}(q^{j-2}z)\right)\right\}:,\nonumber
\\
x_N^+(z)
&=&\sum_{j=1}^N 
:\exp\left(
(b+c)^{j,N}(q^{j-1}z)
+b^{j,N+1}(q^{j-1}z)
-\sum_{l=1}^{j-1}(b_+^{l,N+1}(q^lz)+b_+^{l,N}(q^lz))\right):,
\end{eqnarray}
\begin{eqnarray}
x_i^-(z)&=&
q^{k+N-1}
:\exp\left(a_+^i(q^{\frac{k+N-1}{2}}z)
-b^{i,N+1}(q^{k+N-1}z)-b_+^{i+1,N+1}(q^{k+N-1}z)+b^{i+1,N+1}(q^{k+N}z)
\right):\nonumber
\\
&+&
\frac{1}{(q-q^{-1})z}
\sum_{j=1}^{i-1}
:\exp\left(
a_-^i(q^{-\frac{k+N-1}{2}}z)\right)\times
\\
&&\times
\exp\left((b+c)^{j,i+1}(q^{-k-j}z)
+b_-^{i,n+1}(q^{-k-n}z)-b_-^{i+1,n+1}(q^{-k-n+1}z)
\right)
\nonumber\\
&&\times
\exp\left(
\sum_{l=j+1}^i 
(b_-^{l,i+1}(q^{-k-l+1}z)-b_-^{l,i}(q^{-k-l}z))
+\sum_{l=i+1}^N
(b_-^{i,l}(q^{-k-l}z)-b_-^{i+1,l}(q^{-k-l+1}z))\right)
\times
\nonumber\\
&&
\times
\left\{
\exp\left(-b_-^{j,i}(q^{-k-j}z)
-(b+c)^{j,i}(q^{-k-j+1}z)\right)
-
\exp\left(
-b_+^{j,i}(q^{-k-j}z)
-(b+c)^{j,i}(q^{-k-j-1}z)
\right)
\right\}:
\nonumber\\
&+&\frac{1}{(q-q^{-1})z}
:\left\{
\exp\left(a_-^i(q^{-\frac{k+N-1}{2}}z)
+(b+c)^{i,i+1}(q^{-k-i}z)\right.\right.\nonumber\\
&&\left.+\sum_{l=i+1}^N(b_-^{i,l}(q^{-k-l}z)
-b_-^{i+1,l}(q^{-k-l+1}z))
+b_-^{i,N+1}(q^{-k-N}z)-b_-^{i+1,N+1}(q^{-k-N+1}z)\right)
\nonumber
\\
&-&
\exp\left(a_+^i(q^{\frac{k+N-1}{2}}z)
+(b+c)^{i,i+1}(q^{k+i}z)\right.\nonumber\\
&&\left.\left.
+\sum_{l=i+1}^N(b_+^{i,l}(q^{k+l}z)
-b_+^{i+1,l}(q^{k+l-1}z))
+b_+^{i,N+1}(q^{k+N}z)-b_+^{i+1,N+1}(q^{k+N-1}z)\right)\right\}:
\nonumber\\
&-&\frac{1}{(q-q^{-1})z}
\sum_{j=i+1}^{N-1}
:\exp\left(
a_+^i(q^{\frac{k+N-1}{2}}z)\right)\times
\nonumber\\
&&\times
\exp\left((b+c)^{i,j+1}(q^{k+j}z)
+b_+^{i,N+1}(q^{k+N}z)-b_+^{i+1,N+1}(q^{k+N-1}z)\right)
\times
\nonumber\\
&&\times
\exp\left(\sum_{l=j+1}^N
(b_+^{i,l}(q^{k+l}z)-b_+^{i+1,l}(q^{k+l-1}z))
\right)
\times
\left\{
\exp\left(b_+^{i+1,j+1}(q^{k+j}z)-(b+c)^{i+1,j+1}(q^{k+j+1}z)\right)
\right.\nonumber\\
&&\left.-
\exp\left(b_-^{i+1,j+1}(q^{k+j}z)-(b+c)^{i+1,j+1}(q^{k+j-1}z)\right)
\right\}:.
\nonumber
\end{eqnarray}
\begin{eqnarray}
x_N^-(z)&=&
\frac{1}{(q-q^{-1})z}
\left\{\sum_{j=1}^{N-1}
q^{j-1}:\exp\left(
a_-^N(q^{-\frac{k+N-1}{2}}z)\right)\times
\right.\\
&&
\times 
\exp\left(
-b_+^{j,N+1}(q^{-k-j}z)-b^{j,N+1}(q^{-k-j-1}z)
-\sum_{l=j+1}^{N-1}(b_-^{l,N}(q^{-k-l}z)
+b_-^{l,N+1}(q^{-k-l}z))\right)\times
\nonumber
\\
&&\times
\left\{
\exp\left(-b_+^{j,N}(q^{-k-j}z)-(b+c)^{j,N}(q^{-k-j-1}z)\right)
-
\exp\left(
-b_-^{j,N}(q^{-k-j}z)-(b+c)^{j,N}(q^{-k-j+1}z)\right)
\right\}:\nonumber\\
&+&\left.
q^{N-1}:\left\{
\exp\left(
a_+^N(q^{\frac{k+N-1}{2}}z)-b^{N,N+1}(q^{k+N-1}z)\right)
-
\exp\left(
a_-^N(q^{-\frac{k+N-1}{2}}z)-b^{N,N+1}(q^{-k-N+1}z)\right)
\right\}:
\right\}.
\nonumber
\end{eqnarray}
\begin{eqnarray}
\psi_i^\pm(q^{\pm \frac{k}{2}}z)&=&
\exp\left(
a_\pm^i(q^{\pm \frac{k+N-1}{2}}z)+
\sum_{l=1}^i(b_\pm^{l,i+1}(q^{\pm(l+k-1)}z)-b_\pm^{l,i}
(q^{\pm(l+k)}z)\right)
\\
&\times&
\exp\left(\sum_{l=i+1}^{N}(b_\pm^{i,l}(q^{\pm(k+l)}z)-
b_\pm^{i-1,l}(q^{\pm(k+l-1)}z)
+b_\pm^{i,N+1}(q^{\pm(k+N)}z)-
b_\pm^{i+1,N+1}(q^{\pm(k+N-1)}z)\right),
\nonumber
\\
\psi_N^\pm(q^{\pm \frac{k}{2}}z)
&=&
\exp\left(a_\pm^N(q^{\pm \frac{k+N-1}{2}}z)-
\sum_{l=1}^{N-1}
(b_\pm^{l,N}(q^{\pm (k+l)}z)
+b_\pm^{l,N+1}(q^{\pm (k+l)}z))\right).
\end{eqnarray}
\end{thm}

\subsection{Elliptic Superalgebra $U_{q,p}(\widehat{sl}(N|1))$}

In this section we give
a bosonization of the elliptic superalgebra
$U_{q,p}(\widehat{sl}(N|1))$ for an arbitrary level $k$,
using the dressing deformation \cite{Kojima2}.
Let us introduce the zero-mode operators 
$P_i,Q_i$, $(1\leq i \leq N)$ by
\begin{eqnarray}
~[P_i,Q_j]=-\frac{A_{i,j}}{2}~~(1\leq i,j \leq N),
\end{eqnarray}
where $(A_{i,j})_{1\leq i,j \leq N}$ is the Cartan matrix
of the classical $sl(N|1)$.
In \cite{Kojima2}
the bosonization of the Drinfeld generator
$h_{i,m}$ $(1\leq i \leq N, m \in {\bf Z})$ is given by
\begin{eqnarray}
h_{i,m} &=&
q^{-\frac{k+N-1}{2}|m|}a_m^i+
\sum_{l=1}^i(q^{-(k+l-1)|m|}b_m^{l,i+1}
-q^{-(k+l)|m|}b_m^{l,i})
\\
&&+
\sum_{l=i+1}^N(q^{-(k+l)|m|}b_m^{i,l}-
q^{-(k+l-1)|m|}b_m^{i+1,l})+q^{-(k+N)|m|}b_m^{i,N+1}-
q^{-(k+N-1)|m|}b_m^{i+1,N+1},
\nonumber
\\
h_{N,m}
&=&
q^{-\frac{k+N-1}{2}|m|}a_m^N-
\sum_{l=1}^{N-1}(q^{-(k+l)|m|}b_m^{l,N}+q^{-(k+l)|m|}
b_m^{l,N+1}).
\end{eqnarray}
Let us set the boson $B_{j,m}$ 
$(1\leq j \leq N, m \in {\bf Z}_{\neq 0})$ by
\begin{eqnarray}
B_{j,m}=\left\{
\begin{array}{cc}
\frac{[r^*m]_q}{[rm]_q}h_{j,m}&~~~(m>0),\\
q^{k|m|}h_{j,m}&~~~(m<0).
\end{array}
\right.
\end{eqnarray}

\begin{thm}~~
\cite{Kojima1, Kojima2}~~
The currents 
$E_j(z), F_j(z), H_j^\pm(z)$ 
$(1\leq j \leq N)$ of the elliptic superalgebra
$U_{q,p}(\widehat{sl}(N|1))$ 
for an arbitrary level $k$ are realized by
the bosonic operators as follows.
\begin{eqnarray}
&&E_j(z)=U_j^+(z)x_j^+(z)e^{2Q_j}z^{-\frac{1}{r^*}P_j},\\
&&F_j(z)=x_j^-(z)U_j^-(z)z^{\frac{1}{r}(P_j+h_j)},\\
&&H_j^\pm(z)=H_j(q^{\pm(r-\frac{k}{2})}z),\\
&&H_j(z)=:\exp\left(-\sum_{m\neq 0}
\frac{B_{j,m}}{[r^*m]_q}z^{-m}
\right):e^{2Q_j}z^{-\frac{k}{r r^*}P_j+\frac{1}{r}h_j}.
\end{eqnarray}
Here we have used the dressing operators
$U_j^+(z), U_j^-(z)$ $(1\leq j \leq N)$ given by
\begin{eqnarray}
U_j^+(z)=\exp\left(
\sum_{m>0}\frac{q^{rm}}{[rm]_q}B_{j,-m}z^{m}\right),~
U_j^-(z)=\exp\left(
-\sum_{m>0}\frac{q^{r^* m}}{[rm]_q}B_{j,m}z^{-m}\right).
\end{eqnarray}
\end{thm}

~\\

\section{$q$-Wakimoto Realization}

In this section we introduce
the $q$-Wakimoto realization
by the $\xi$-$\eta$ system.
We introduce the vacuum state $|0\rangle$ of 
the boson Fock space by
\begin{eqnarray}
a_m^i|0\rangle=b_m^{i,j}|0\rangle
=c_m^{i,j}|0 \rangle=0~~(m \geq 0).
\end{eqnarray}
For complex numbers $p_a^i \in {\bf C}$ $(1\leq i \leq N)$,
$p_b^{i,j} \in {\bf C}$ $(1\leq i<j \leq N+1)$,
$p_c^{i,j} \in {\bf C}$ $(1\leq i<j \leq N)$,
we set
\begin{eqnarray}
&&|p_a, p_b, p_c \rangle
\\
&=&\exp\left(
\sum_{i,j=1}^N
\frac{{\rm Min}(i,j)(N-1-{\rm Max}(i,j))}{(N-1)(k+N-1)}
p_a^i Q_a^j
-\sum_{1\leq i<j \leq N+1}p_b^{i,j}Q_b^{i,j}
+\sum_{1\leq i<j \leq N}p_c^{i,j}Q_c^{i,j}
\right)|0\rangle.\nonumber
\end{eqnarray}
It satisfies
\begin{eqnarray}
a_0^i|p_a,p_b,p_c\rangle=p_a^i |p_a,p_b,p_c\rangle,~
b_0^{i,j}|p_a,p_b,p_c\rangle=p_b^{i,j} |p_a,p_b,p_c\rangle,~
c_0^{i,j}|p_a,p_b,p_c\rangle=p_c^{i,j} |p_a,p_b,p_c\rangle.
\end{eqnarray}
The boson Fock space $F(p_a,p_b,p_c)$
is generated by
the bosons $a_m^i, b_m^{i,j}, c_m^{i,j}$
on the vector $|p_a,p_b,p_c\rangle$.
We set $U_q(\widehat{sl}(N|1))$-module $F(p_a)$ by
\begin{eqnarray}
F(p_a)=
\bigoplus
_{
p_b^{i,j}=-p_c^{i,j} \in {\bf Z}~(1\leq i<j \leq N)
\atop{
p_b^{i,N+1} \in {\bf Z}~(1\leq i \leq N)
}}F(p_a,p_b,p_c).
\end{eqnarray}
We have imposed the restriction
$p_b^{i,j}=-p_c^{i,j} \in {\bf Z}$,
because 
the $x_{i,m}^\pm$ change 
$Q_b^{i,j}+Q_c^{i,j}$.
The module $F(p_a)$
is not irreducible representation.
For instance, the irreducible highest weight
module $V(\lambda)$ for 
$U_q(\widehat{sl}(2))$
was constructed from the similar space as $F(p_a)$
by two steps; the first step is
the construction of the $q$-Wakimoto
realization by the $\xi$-$\eta$ system, and
the second step is the resolution
by the Felder complex \cite{Konno}.
In this paper we study the $\xi$-$\eta$ system
and introduce the $q$-Wakimoto realization 
for $U_q(\widehat{sl}(N|1))$.
For $1\leq i< j \leq N$ we introduce
\begin{eqnarray}
\eta^{i,j}(z)=\sum_{m \in {\bf Z}}\eta_{m}^{i,j}
z^{-m-1}=:e^{c^{i,j}(z)}:,~~
\xi^{i,j}(z)=\sum_{m \in {\bf Z}}\xi_{m}^{i,j}z^{-m}=:e^{-c^{i,j}(z)}:.
\end{eqnarray}
The Fourier components $\eta_m^{i,j}
=\oint \frac{dz}{2\pi \sqrt{-1}}z^m \eta^{i,j}(z)$,
$\xi_m^{i,j}=
\oint \frac{dz}{2\pi \sqrt{-1}}z^{m-1}\xi^{i,j}(z)$ $(m \in {\bf Z})$
are well defined on the space ${F}(p_a)$.
They satisfy
\begin{eqnarray}
&&
\{\eta_m^{i,j},\xi_n^{i,j}\}=\delta_{m+n,0},~
\{\eta_m^{i,j},\eta_n^{i,j}\}=\{\xi_m^{i,j},\xi_n^{i,j}\}=0~~~
(1\leq i<j \leq N),\\
&&~[\eta_m^{i,j},\xi_n^{i',j'}]=[\eta_m^{i,j},\eta_n^{i',j'}]
=[\xi_m^{i,j},\xi_n^{i',j'}]=0~~~(i,j)\neq (i',j').
\end{eqnarray}
We focus our attention on the operators 
$\eta_0^{i,j}, \xi_0^{i,j}$ satisfying
$(\eta_0^{i,j})^2=0$, $(\xi_0^{i,j})^2=0$ and
${\rm Im} (\eta_0^{i,j})={\rm Ker} (\eta_0^{i,j})$,
${\rm Im} (\xi_0^{i,j})={\rm Ker} (\xi_0^{i,j})$.
The products $\eta_0^{i,j} \xi_0^{i,j}$ and 
$\xi_0^{i,j} \eta_0^{i,j}$ are 
the projection operators
\begin{eqnarray}
&&\eta_0^{i,j}\xi_0^{i,j}+\xi_0^{i,j}\eta_0^{i,j}=1.
\end{eqnarray}
We have a direct sum decomposition.
\begin{eqnarray}
&&F(p_a)=
\eta_0^{i,j}\xi_0^{i,j}F(p_a) \oplus 
\xi_0^{i,j}\eta_0^{i,j}F(p_a),
\\
&&{\rm Ker} (\eta_0^{i,j})=
\eta_0^{i,j}\xi_0^{i,j}F(p_a),~
{\rm Coker} (\eta_0^{i,j})=
\xi_0^{i,j}\eta_0^{i,j}F(p_a).
\end{eqnarray}
\begin{dfn}~~~
We introduce the subspace ${\cal F}(p_a)$ 
that we call the $q$-Wakimoto realization.
\begin{eqnarray}
{\cal F}(p_a)
=\left(\prod_{1\leq i<j \leq N}
{\eta_0^{i,j} \xi_0^{i,j}}\right)
F(p_a)
=\bigcap_{1 \leq i<j \leq N}{\rm Ker}(\eta_0^{i,j}),
\end{eqnarray}
\end{dfn}
The dressing operators $U^\pm_{i}(z)$ and the zero-mode operators 
$P_i, Q_i$ commute with $\eta_0^{i',j'}$.
The bosonizations
commute with
the operators $\eta_0^{i',j'}$,$\xi_0^{i',j'}$ 
up to sign $\pm$.
\begin{prop}~~~
The subspace ${\cal F}(p_a)$ is both $U_q(\widehat{sl}(N|1))$ 
and $U_{q,p}(\widehat{sl}(N|1))$ module.
\end{prop}
Let $\bar{\alpha}_i$, $\bar{\Lambda}_i$, 
$(1\leq i \leq N)$ and $(\cdot|\cdot)$
be the simple roots, the fundamental weights,
and the symmetric bilinear norm ;
$(\bar{\alpha}_i,\bar{\alpha}_j)=A_{i,j}$, 
$(\bar{\alpha}_i,\bar{\Lambda}_j)=\delta_{i,j}$.
It is expected that we have
the irreducible highest weight module $V(\lambda)$
with the highest weight $\lambda$,
whose classical part 
$\bar{\lambda}=\sum_{j=1}^N p_a^i \bar{\Lambda}_i$,
by the Felder complex 
of the $q$-Wakimoto realization.
We would like to report this problem for
$U_q(\widehat{sl}(N))$ and $U_q(\widehat{sl}(N|1))$
in the future publication.


\subsection*{Acknowledgements}~~
This work is supported by the Grant-in-Aid for
Scientific Research {\bf C} (21540228)
from Japan Society for Promotion of Science. 
The author would like to thank
the organizing committee of
the 9-th International Workshop
"Lie Theory and its application in Physics"
for an invitation to Bulgaria.

\begin{appendix}

\end{appendix}

\end{document}